\documentclass[12pt,english]{elsarticle}
\usepackage{mathpazo}
\usepackage[T1]{fontenc}
\usepackage[latin9]{inputenc}
\usepackage{geometry}
\usepackage{fancyhdr}
\usepackage{color}
\usepackage{float}
\usepackage{textcomp}
\usepackage{amsmath}
\usepackage{amssymb}
\usepackage{graphicx}
\usepackage{esint}

\makeatletter

\providecommand{\tabularnewline}{\\}

\@ifundefined{date}{}{\date{}}

\usepackage{babel}
\newcommand{\ve}[1]{\ensuremath{\mbox{\boldmath$#1$}}}
\newcommand{\vecv}{\ve {v}}
\newcommand{\vecx}{\ve {x}}

\makeatother

\usepackage{babel}
\begin{document}
\begin{frontmatter}{}

\title{On the emission of ultra fine particles from municipal solid waste
(MSW) incinerators}

\author{M. W. Reeks}

\address{Dept. of Aeronautics, Imperial College,\\
 Exhibition Rd, London SW7 2AZ\\
 email: mike.reeks@yahoo.com}
\begin{abstract}
\noindent Nationally approved emission factors of mass versus particle
size for particulate emissions from UK MSW incinerators when converted
to particle number versus size, indicate that nearly all $(>90\%$)
of the emitted particles are ultra fine particles (ufps) < .1 micron
in size. A similar result is true also of US MSW incinerators emissions.
This would imply that the bag/fiber filters used for the removal of
particles produced in the incineration process have a very low efficiency
for the removal of ufps. This result is at variance with recent assertions
that bag filters have a high removal efficiency for ufps. An analysis
of fiber filter retention based on the fundamental mechanisms for
the deposition of small particles to single filter fibers and their
dependence on particle size, shows that whilst the removal efficiency
is 100 \% for particles <\textcompwordmark < . 1 micron, there is
a minimum of the filter retention efficiency in the region 0.05 to
0.5 microns where the concentration of the ufps is most likely to
be greatest. In some cases depending on the flow and particle size,
the filter efficiency is as low as 5\% compared to almost 100\% retention
efficiency for particles $>$ 1 micron (within the inertial impaction
range of particles). It is believed this explains the very high release
rates $\sim10^{14}$ particles/s from these incinerators. 
\end{abstract}
\end{frontmatter}{}

\section{Introduction\label{sec:Introduction}}

In the UK, incineration is rapidly replacing land fill sites for the
removal and disposal of municipal solid waste (MSW) (Nixon et al.
2013; Bowden 2019). Whilst MSW incineration has the additional advantage
that it can be used as a viable source of heat and power, there is
however a very real health concern associated with the inhalation
of toxic gases and aerosols produced during the incineration/combustion
process, adding to the pollution from traffic/vehicle emissions and
increasing the overall health risk to the public (Oberdorster 2001)
(especially to those in the immediate vicinity of an incinerator (Ghosha
et al. 2019. Furthermore recent analyses of MSW (Shang et al. 2019)
have shown that the aerosol released is far more toxic than that arising
from traffic/vehicle exhausts. Of particular concern is the release
of ultra fine particles (ufps) $\leq$ .1 micron in size since their
inhalation and deposition in the human respiratory tract gives rise
to chronic asthma (in the upper airways) and cardio vascular diseases
when they are absorbed into the bloodstream via the alveolar deeper
recesses of the human lungs.\\
 \\
 There are currently over 40 incinerators distributed around the UK.
Their particulate emission rates are currently regulated and controlled
by limitations on the annual total mass released together with stipulated
emission factors (EFs) (fractions of the total annual mass released)
associated with the various size ranges that make up the total mass
released. Much of the abatement and control of these emissions relies
on the effective use of filters at the end of the chain of combustion
and chemical treatment prior to release from the incinerator stack
to the atmosphere. Of particular concern is the filter removal efficiency
for ufps since they represent the greatest health risk. This seriously
questions the value of using a size distribution for particulate emissions
based on particle mass rather than particle numbers, which is a much
more sensitive measure for ufp related emissions. In particular it
is shown in the next section that in converting mass to particle number
for quoted annual incinerator releases, $>$90 \% of the number of
particles are ufps. It suggests that the incinerator filters whilst
highly efficient in removing the larger size particles (and largely
responsible for the total mass released ) are ineffective in removing
ufps.

\begin{table}[H]
\centering{}\includegraphics[scale=0.5]{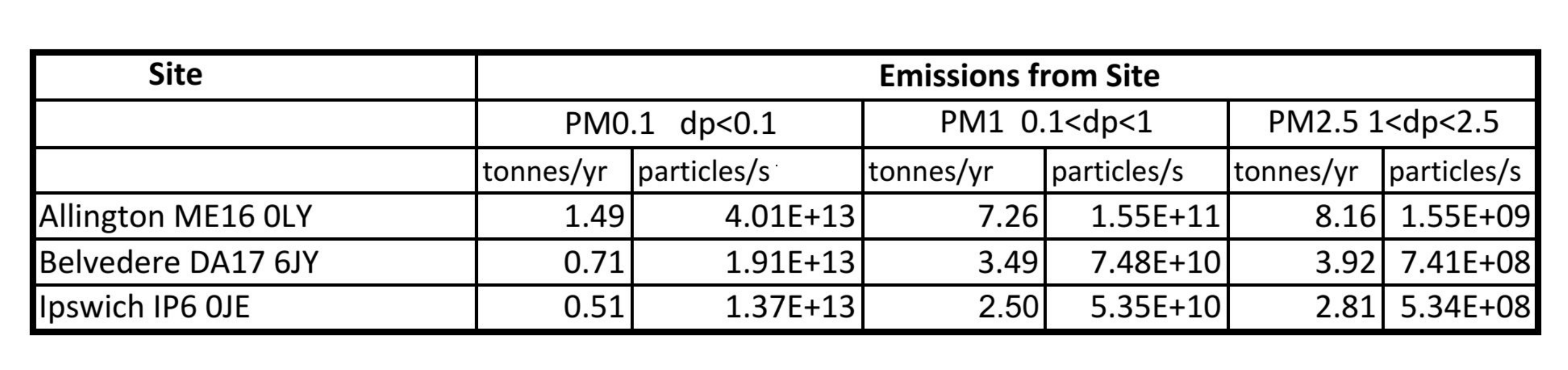}\caption{\label{tab:annual-particulate-mass-1}Annual particulate mass emitted
converted to particle numbers/sec for 3 typical UK incinerators \protect
\protect \\
 Data obtained from the National Atmospheric Emissions Inventory (NAEI
data base 2016) }
\end{table}

Most of the filters used in current incinerators are bag or fiber
filters. A recent review by Jones and Harrison (2016) claims that
bag filters can be as effective at removing ufps as the larger size
range of inertial particles$\geq$ 1 micron. They point out that ufp
removal efficiencies are far greater than one would expect from a
simple sieving mechanism based on sieve size and that furthermore
they have almost 100\% removal efficiency over the entire range of
particle sizes. The main purpose of this paper is to examine the validity
of this assertion, showing that in particular it is inconsistent with
(a) the quoted annual mass release rates rate when converted to particle
numbers and (b) the calculated removal efficiencies based on the fundamental
mechanisms for the deposition and retention of small particles within
a fiber filter. We show that fiber filters behave in a very similar
way to that of the human respiratory tract where inhaled ufps unlike
the larger size PM10 particles are not deposited in the upper airways
of the lungs but penetrate down into the deeper alveolar regions.
In particular there exists a range of submicron particles $\leq.1$
microns in size for which the fiber removal efficiencies are very
small. In this paper we examine the properties of this range of particles
which have the maximum penetration in fiber filters and show that
the resulting filter performance over the entire range of particles
size is consistent with the observed particulate emission from an
incinerator.

\section{Conversion of mass to particle number for incinerator releases }

\noindent Details are given here of the conversion of the particle
mass to particle numbers and how that leads to the conclusion that
most of the particles released from UK incinerators are ufps. It highlights
the importance of using particle number versus size distributions
rather than particle mass versus size as a more appropriate measure
when dealing with ufp related emissions. \\
 Table \ref{tab:annual-particulate-mass-1} gives the measured annual
mass releases of particulate in the various PM size ranges. Figures
for most of the 40 incinerators are available\textcolor{blue}{{}
(NAEI 2016).} The three incinerators referred to in Table \ref{tab:annual-particulate-mass-1}
are just typical examples. Conversion of mass to particle number in
the size ranges shown in Table \ref{tab:annual-particulate-mass-1}
are based on the maximum size in a given size range assuming that
all particles are spherical with a particle diameter given by the
maximum particle size in the particular size range. So the number
of particles $N_{Dp}$ associated with the total particle mass $M_{Dp}$
in a particular size range (band) is given simply by 
\begin{equation}
N_{Dp}=M_{Dp}/\left(\pi D_{p}^{3}\rho_{p}/6\right)\label{eq:minimum particle number}
\end{equation}
where $\rho_{p}$ is the material density of the incinerated MSW which
we have taken to be the material density of carbon of 2.25$\times10^{3}\;kg/m^{3}$.
$D_{p}$ is the PM particle size and by definition the maximum value
of the particle size in the given size range taken from the annual
mass releases of PM.1, PM1, PM2.5 particles. So the particle numbers
/ sec given in Table \ref{tab:annual-particulate-mass-1} are the
minimum numbers of particles released / sec based on the assumption
of uniform material density of the MSW, and assuming a discrete particle
size given by the maximum size in each size range. The actual number
is likely to be significantly higher depending upon the number size
distribution in each size range and is given by the ratio of the volume
average $\overline{v}_{D_{p}}$ for PM $D_{p}$ particles to the volume
$v_{Dp}$ of the particle of size $D_{p}$ assuming the volume $v_{p}$
for a particle of size $d_{p}\leq D_{p}$ to be that of an equivalent
volume sphere of size $d_{p}$, i.e. $v_{p}=\pi d_{p}^{3}/6$.

\begin{figure}[H]
\centering{}\includegraphics[scale=0.27]{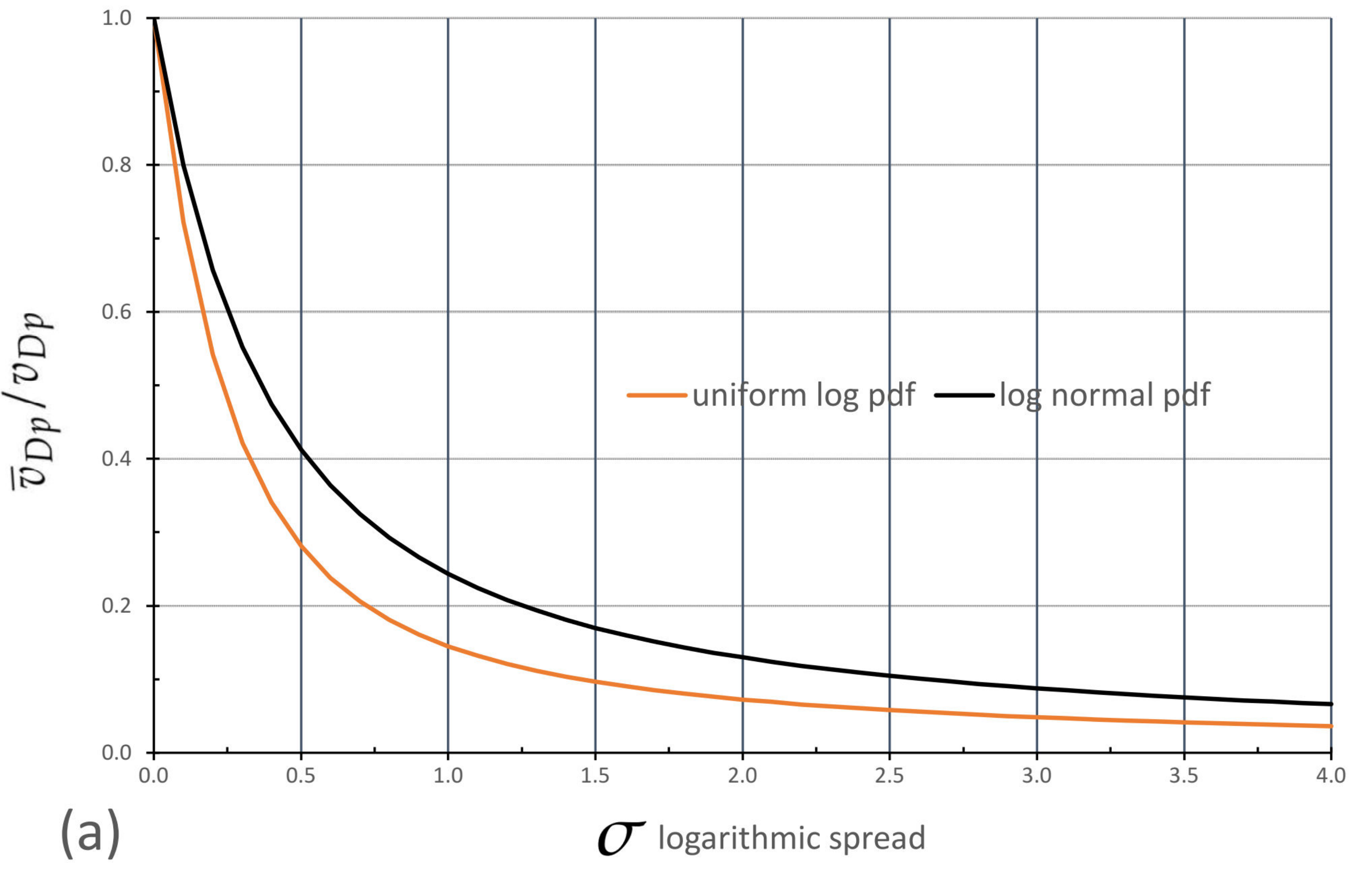}\includegraphics[scale=0.38]{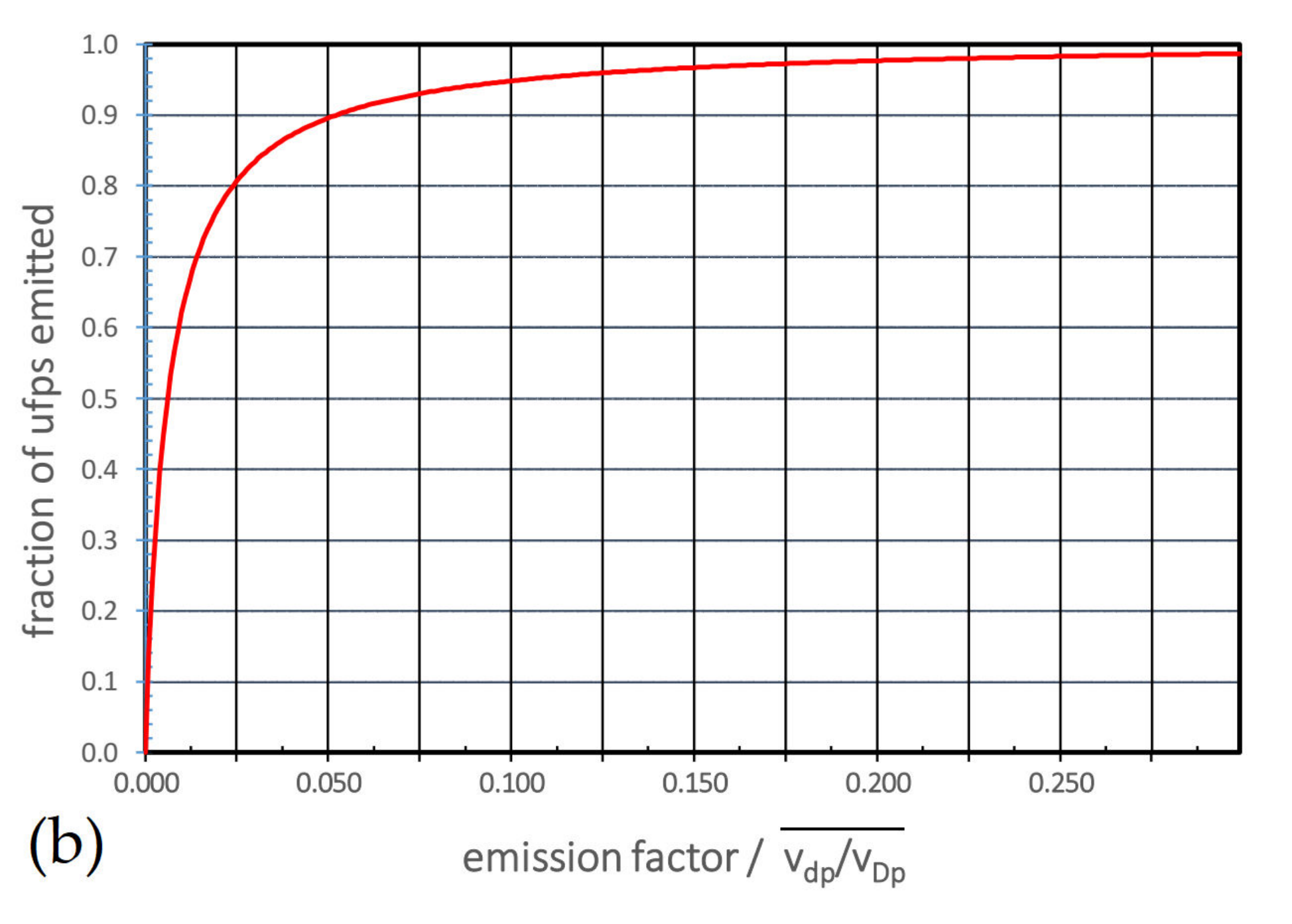}\caption{\label{(a) mean vol sphere}(a) Mean equivalent volume sphere of particle
size distribution versus logarithmic spread \protect \protect \\
 (b) Fraction of ufps of total total number of particles emitted versus
emission factor}
\end{figure}

As as an illustration, values of the ratio $\overline{v_{d_{p}}/v_{D_{p}}}=\overline{\left(d_{p}/D_{p}\right)^{3}}$
are plotted in Fig. \ref{(a) mean vol sphere}(a) as a function of
the spread $\sigma$ of the number size distribution for 2 particle
number-size distributions as examples: $(a)$ a uniform distribution
$P(z)$, $z=log_{10}\left(d_{p}/D_{p}\right),\;-\sigma\leq z\leq0$
with $P(z)=\sigma^{-1},$$(b)$ a log-normal distribution $P(z)=\sqrt{2/\pi}exp(-z^{2}/2\sigma^{2})$,
$z=ln(dp/D_{p}),-\infty<z<0$. We note that for $\sigma=0$, $\overline{v_{d_{p}}/v_{D_{p}}}=1$
corresponding to the discrete distribution $\delta$($ln(z)$). For
both distributions, for $\sigma\gg1$ $\overline{v_{d_{p}}/v_{D_{p}}}\rightarrow\sigma^{-1}$,
See Table \ref{tab:-Logarithmic-distributions} for the precise asymptotic
values for $\sigma\gg1$\\

\noindent 
\begin{table}[H]
\noindent \begin{centering}
\begin{tabular}{|c|c|c|}
\hline 
Distribution $P(z)$  & $z\gg1$  & $\text{\ensuremath{\overline{\left(v_{dp}/v_{D_{p}}\right)}}}$\tabularnewline
\hline 
\hline 
uniform $\sigma^{-1}$  & $log_{10}(d_{p}/D_{p}),-\sigma\leq z\leq0$  & $\left(3\sigma ln(10)\right)^{-1}$\tabularnewline
\hline 
lognormal $\sqrt{2/\pi}exp(-z^{2}/2\sigma^{2})$  & $ln(d_{p}/D_{p})-\infty\leq z\leq0$  & $\sqrt{2/\pi}/3\sigma$\tabularnewline
\hline 
\end{tabular}
\par\end{centering}
\caption{\label{tab:-Logarithmic-distributions}Logarithmic distributions and
asymptotic forms for the spread $\sigma\rightarrow\infty$, used in
the evaluation of $\overline{v_{dp}/v_{D_{p}}}$, and the plots given
in Fig. \ref{(a) mean vol sphere}(b). So to calculate the particle
numbers / sec for each of the 2 distributions, you divide the number
$N_{Dp}$ given by the formula in Eq.(\ref{eq:minimum particle number})
for the minimum number of particles by the value $\overline{v_{d_{p}}/v_{D_{p}}}$
in Figure \ref{(a) mean vol sphere} (a) for a particular value of
$\sigma$.}
\end{table}

It is important to point out that only the total annual mass released
is a measured quantity for each incinerator. The actual amount for
each particular size range / limit is given by an emission factor
(EF) which is based on the average of numerous measurements of the
mass distribution of sampled concentrations of airborne emissions.
Their recommended values are applicable to all UK incinerator emissions.
The fraction of the total number of particles in each range is determined
only by the emission factor $E_{f}$ and the conversion of mass to
number. Fig. \ref{(a) mean vol sphere} (b) shows the dependence of
the number fraction of ufps for $d_{p}\leq0.1$ microns on the ratio
$E_{f}/\overline{v_{d_{p}}}$ . It shows an initial sharp rise in
the ufp number fraction for small values of $E_{f}/\overline{v_{d_{p}}/v_{D_{p}}}$
and flattening out for values $\gtrsim0.05,$ where the ufp $\gtrsim0.9$.
In this range the ufp fraction is very insensitive to changes in the
value the emission factor $E_{f}/\overline{v_{d_{p}}/v_{D_{p}}}$.
Table \ref{tab 3 revised} gives values for the number fraction for
each size range of particles $\left[d_{p}\leq0.1;\:0.1\leq d_{p}\leq1,\:1\leq d_{p}\leq2.5\right]$
based on the NAEI figures for the emission factors and also those
of the MSW US incinerators quoted in the recent CEDA data base (Harrison
et al. 2019). Assuming e.g. an average value of $\overline{v_{p}/v_{D_{p}}}$
$=1$ as was assumed in Table \ref{tab:annual-particulate-mass-1}
whilst giving a minimum value for the number of particles in each
size range, this value gives an over estimate the of the ufp fraction
of the total number of particles emitted. So to be more realistic
in Table \ref{tab 3 revised} we have used values for $\overline{v_{p}/v_{D_{p}}}$
based on a uniform logarithmic spread given in Fig. \ref{(a) mean vol sphere}.
In particular for particle diameters $d_{p}$, $0.1\leq d_{p}\leq1$
$\overline{v_{p}/v_{D_{p}}}=0.145$ ;$1\leq d_{p}\leq2.5$ $\overline{v_{p}/v_{D_{p}}}=0.34$.
Table \ref{tab 3 revised} gives values for the ufp fraction for the
case when all ufps have a diameter of $0.1$ microns and when $\overline{v_{p}/v_{D_{p}}}$=
0.145 corresponding to the case when particles sizes are uniformly
distributed logarithmically for $0.01\leq d_{p}\leq0.1$. There is
little difference between the NAEI and the USEPA results, the fraction
of the number of ufps of the total number of particles from either
source is almost 100\% \\

\begin{table}[H]
\centering{}\includegraphics[scale=0.45]{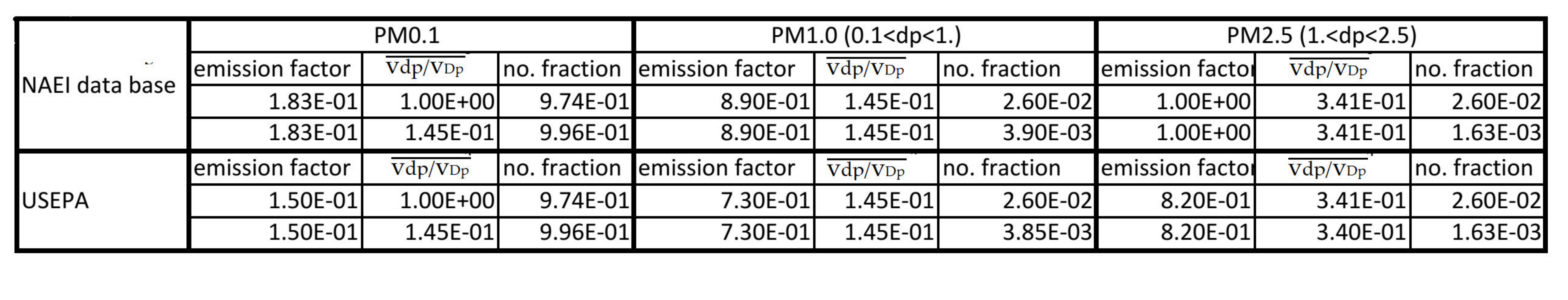}\caption{\label{tab 3 revised}\% age of total number of particles in each
size range, based on emission factors obtained from the NAEI data
base used for emissions from UK incinerators and US EPA for MSW (Harrison
ed. 2017)}
\end{table}

\section{Filtration Mechanisms and removal efficiencies\label{sec:Filtration-Mechanisms}}

Filtration especially filtration by fibers is a much studied area
of research both experimentally and theoretically. There is a huge
literature on the subject and a number of useful and informative reviews
have been published that can usefully be referred to when considering
the filtration of ultrafine particles (ufps) from incinerator emissions
(Wang 2013; Yamada et al. 2011; Stechkina and Fuchs 1966; Harrington
et al. 2007 ). In considering the removal efficiency of bag filters
we examine here the influence of the various mechanisms responsible
for the collection of small particles by cylindrical fibers in a flowing
gas suspension at low Reynolds number $Re$ viscous flow:

\begin{figure}[H]
\begin{centering}
\includegraphics[scale=0.25]{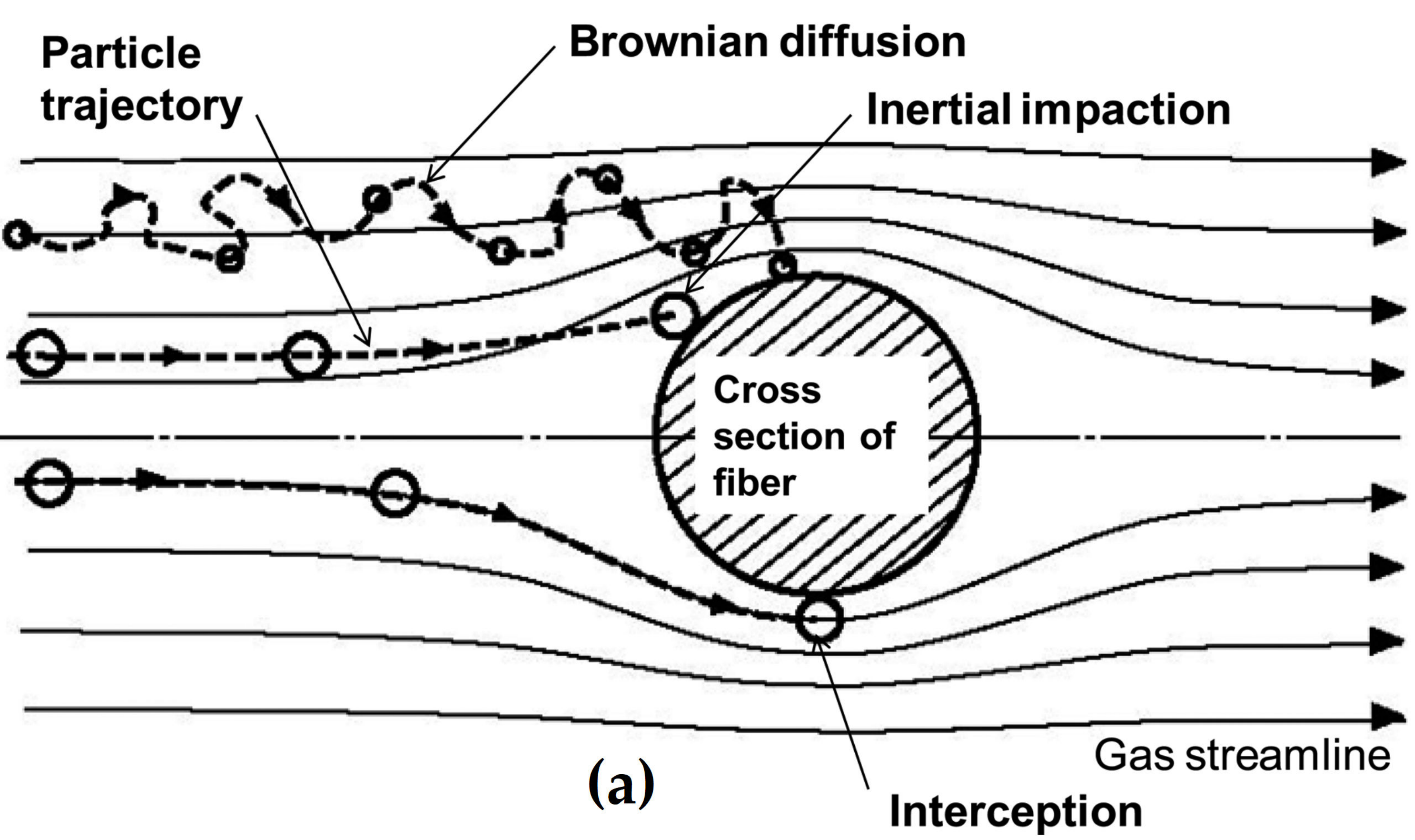}\includegraphics[scale=0.28]{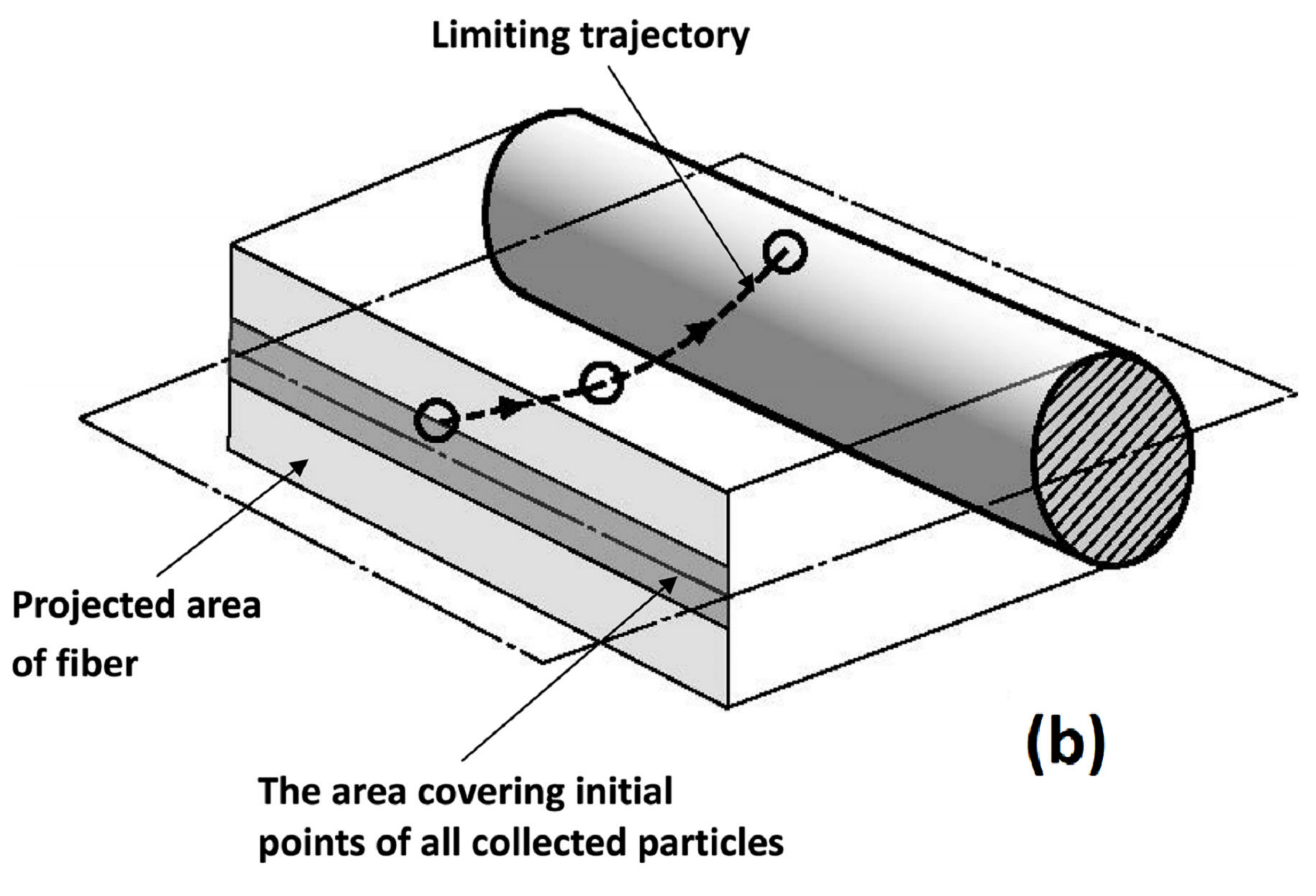}\\
\par\end{centering}
\centering{}\caption{\label{fig:collection mechanisms}(a) Particle collection by inertial
impaction, interception, and convective Brownian diffusion; (b) Definition
of single fiber collection efficiency. Figures taken from Wang C-sen
and Otani (2013) }
\end{figure}

How in particular the penetration/ collection efficiency of the filter
depends upon the flow and particle size down to the molecular size?
So in section \ref{sec:Filtration-Mechanisms} we briefly describe
the various mechanisms by which a suspended particle can be deposited
to the surfaces of a cylindrical fiber in the cross flow of a gas
particle suspension and how as a result, this deposition depends on
2 numbers, the Peclet number $Pe$ that determines the contribution
of particle diffusion over convection and the Stokes number $St$
that measures the extent that particles are projected out of the flow
onto the surfaces of the cylinder as a result of their inertia. In
so doing we consider the various empirical formulae reported in the
literature that have been used to accurately determine the collection
efficiency of a cylindrical fiber and how this can be used to calculate
the overall penetration of a filter of a given fiber density, volume
fraction and thickness. Then in section \ref{sec:Calculation-of-penetration}
using these formulae for the removal efficiencies for single fibers,
we show the results of our calculations of filter penetration for
filters of a given fiber density and thickness as a function of particles
size and filter flow. They demonstrate a significant variation of
penetration and the existence of a maximum penetration that occurs
in the range PM.1 range $(<.1microns)$.

In both fiber filters and in the bifurcating airways of the human
lungs, airborne particles are deposited on to surfaces exposed to
the flow by three mechanisms: interception, impaction and Brownian
diffusion each of which depend critically on particles size and flow
rate. They are illustrated in Figure \ref{fig:collection mechanisms}
(a) depicting the cross flow normal to the axis of an isolated fiber
and showing the way a suspended particles in the flow would deposit
on the fiber surface by these individual mechanisms. With reference
to Figure \ref{fig:collection mechanisms}(b), the collection efficiency
$\varepsilon_{p}$ of the fiber is defined as the ratio of the number
of particles striking the surface of the fiber to the number that
would strike it if the streamlines were not diverted by it. So 
\begin{equation}
\varepsilon_{p}=\frac{\Phi}{n_{0}U_{0}d_{f}}\label{eq:collection efficiency of a single fibre}
\end{equation}
where $\Phi$ is the number of particles deposited per unit length
of fiber per unit time, $n_{0}$ is the concentration of particles
upstream where the flow rate $U_{0}$ is unperturbed by the fiber
and $d_{f}$ is the fiber diameter. Assuming that all particles adhere
on contact, this is the ratio of the volume of gas filtered by the
fiber to the volume of gas that would pass through the projected area
of the fire normal to the flow in the absence of the fiber. In the
absence of particle diffusion, the filtered and non filtered volume
(which flows round the cylinder) is separated by a limiting trajectory
shown in Figure \ref{fig:collection mechanisms}(b) located a vertical
distance $y$ from the centre of the fiber upstream from the fiber
where the flow is unperturbed by the fiber.

In the case of interception, particles follow the flow streamlines,
so the limiting trajectory is a streamline passing a distance $r_{p}$
from the surface of the fiber. In the case of impaction, particles
through their inertia are projected out of the flow onto the surface
of the fiber with a finite impact velocity that depends upon the particle
inertia. So $\varepsilon_{p}$, defined as the ratio of the filtered
volume flow rate to the flow rate in the absence of the fiber, is
given by $\varepsilon_{p}=y/r_{f}$ where $r_{f}$ is the radius of
the cylindrical fiber. In the case of \emph{inertial} impaction, the
filtered volumes is necessarily greater than that for interception
alone, with $\varepsilon_{p}$ $\rightarrow1$ as the particle inertia
increases.

Inertial impaction of a particle is related to the flow field around
the particle, which is characterized in terms of the particle Reynolds
number $Re_{p}(=d_{p}U_{p}/\nu_{f}$), where $d_{p}$ is the particle
diameter, $U_{p}$ is the particle velocity relative to the gas flow,
and $\nu_{f}$ is the kinematic viscosity of the gas flow. For $Re_{p}<1$,
the governing parameter for inertial impaction is the Stokes number
$St$ defined as the ratio of the particle response $\tau_{p}$ to
changes in flow. $\tau_{p}$ is the Stokes relaxation time and the
time scale for changes in flow $\Delta L/U_{f}$, where $\Delta L$
is the length scale for changes in the flow and $U_{f}$ the flow
rate. In the case of spherical particle of diameter $d_{p}$ of material
density $\rho_{p}$ in fiber flow $U_{f}$ for a fiber of radius $r_{f}$,
$\Delta L\sim r_{f}$ and

\begin{equation}
St=d_{p}^{2}\rho_{p}U_{f}C_{s}/9\mu_{f}d_{f}\label{eq:Stokes number}
\end{equation}
where $C_{s}$ is the slip factor accounting for non continuum flow
which depends on the Knudsen number $Kn_{p}=\lambda_{g}/d_{p}$. \\

\begin{equation}
C_{s}=1+Kn_{p}\left\{ 2.33+0.966\,exp\left(-0.4985/Kn_{p}\right)\right\} \label{eq:Cunningham slip factor}
\end{equation}
where $\lambda_{g}$is the mean fee path of the gas molecules (Stechkina
and Fuchs 1966) .

The total collection efficiency of the fiber depends up on the fiber
density and the surface area of the fiber per unit volume of the fiber
filter as well as the collection efficiency $\varepsilon_{p}$ of
an individual fiber. Suppose the fibers occupies a volume fraction
$\alpha_{f}$ in the filter, then the surface to volume ratio of the
fiber $\lambda_{sv}=\text{\ensuremath{4\alpha_{f}}}/\left(\pi d_{f}\right)$,
so that the penetration $p$ (the fraction of particles transmitted
/escaping from the filter) is given by \textcolor{black}{{} 
\begin{equation}
p=\text{\ensuremath{exp\left(-\lambda_{sv}\varepsilon_{p}L_{f}/(1-\alpha_{f})\right)}=\ensuremath{exp}\ensuremath{\left(-4\alpha\ensuremath{_{f}L_{f}}\varepsilon_{p}/\pi\ensuremath{d_{f}(1-\alpha_{f})}\right)}}\label{eq:penetration}
\end{equation}
where $L_{f}$ is the filter thickness, and $d_{f}$ is the fiber
diameter. The factor $(1-\alpha_{f})$ accounts for the fact that
the flow velocity in the filter is a factor $1/(1-\alpha_{f})$ of
that at the face of the filter. }\\
 \textcolor{black}{{} The validity of the formula in Eq. \ref{eq:penetration}
is based on various assumptions : (1) the volume fraction $\alpha_{f}$
is constant within the filter volume, (2) all fibers are straight
and of uniform diameter, (3) all fibers are normal to the flow direction,
and (4) the single fiber efficiency is constant along the depth of
the filter. In practice all of these assumptions will be approximately
satisfied to a lesser and or greater degree . If $\alpha_{f}$ is
non-uniform or not all fibers are normal to the flow direction, the
collection efficiency $\varepsilon_{p}$ $\le$ the value given by
the formula in Eq.\ref{eq:penetration} . }\\
 \textcolor{black}{{} }\\
 \textcolor{black}{{} Theoretical calculations have been made for
single fiber efficiencies based on individual capture mechanisms and
interception acting together with another mechanism. In calculating
the total collection efficiency arising from all mechanisms acting
simultaneously, it is generally assumed that the efficiencies due
to individual mechanisms are additive. This overestimates the overall
efficiency, because the capture of a particle might be counted more
than once. For the cases in which one capture mechanism dominates,
it appears that the efficiency due to that mechanism alone gives a
better value for the overall efficiency than does the sum of all individual
mechanism efficiencies. See Drossinos and Reeks (2005) for the general
approach for dealing with inertal and diffusional transport acting
simultaneously}

\subsection*{\textcolor{black}{Flow round a single fiber }}

\textcolor{black}{Due to flow interference between neighboring fibers,
the flow around a fiber filter will differ from that around an isolated
fiber. Approximations for the flow in these circumstances have been
derived which can be used reliably to solve the particle equation
of motion for convective diffusion. CFD has aslo been used obtain
numerical solutions of the flow field. The flow field considered by
Kuwabara (1959) was for viscous (low Reynolds number) flow in an array
of randomly distributed parallel circular cylinders placed normal
to the flow. According to the solution, the effect of flow interference
can be characterized and incorporated into the flow solution by the
Kuwabara hydrodynamic factor $Ku$, 
\begin{equation}
Ku=-\frac{ln\alpha_{f}}{2}-\frac{3}{4}+\alpha_{f}-\frac{\alpha_{f}^{2}}{4}\label{eq:Kuwabra number}
\end{equation}
The flow field is valid when $Kn$ $\rightarrow0$ (the continuum
flow range). For relatively small slip flow Pich (1966) gave an expression
based on the Kuwabara model, see also Wang C-sen and Otani (2013)
.}

\subsection*{\textcolor{black}{Inertial Impaction}}

\textcolor{black}{For small Stokes number $St$, the formula in Eq.(\ref{eq:effciency (Stk,R, volfrac)}),
(Stechkina and Fuchs 1966) can be used to calculate the single fiber
efficiency $\varepsilon_{Imp}$ due to inertial impaction and }

\textcolor{black}{{} 
\begin{equation}
\varepsilon_{Imp}=\frac{St}{2Ku^{2}}\left[\left(29.6-28\alpha^{9.62}\right)R^{2}-27.5R^{2.8}\right]\label{eq:effciency (Stk,R, volfrac)}
\end{equation}
where $R$ is the interception ratio $\left(d_{p}/d_{f}\right)$.
The formula is accurate for $0.01<R<0.4$ and $0.0035<\alpha<0.111$.
However it does not take into account the influence of slip flow at
the gas\textminus fiber interface. See the comparison of this formula
with the empirical formula due to Landahl and Hermann (1949) based
on experimental measurements for $Re=10$ ,namely 
\begin{equation}
\varepsilon_{Imp}=\frac{St^{3}}{St^{3}+0.77St{}^{2}+0.22}\label{eq:efficiency (Stk)}
\end{equation}
A number of authors have considered a critical Stokes number $St_{cr}$
\begin{equation}
St<St_{cr}\quad\varepsilon_{I}=0\label{eq:Stk_cr}
\end{equation}
which means that particles must have a certain minimum inertia before
they will deposit given by the value of $St_{cr}.$ It reflects the
fact that the curve of $\varepsilon_{I}=\varepsilon_{I}(St)$ does
not intersect the origin but cuts the $St$ - axis at the point $St=St_{cr}.$
It reflects the fact that there are other mechanisms involved in the
deposition process so that for $St<St_{cr}$, the collection efficiency
is not zero. See Wang and Otani 2013 for more details on its value
for both potential and viscous flow.}

\subsection*{Diffusion convection and interception}

The deposition by gradient diffusion and convection is the solution
of the convection diffusion equation which for a compressible flow
can be written as 
\begin{equation}
\partial n(\ve x,t)/\partial t+\ve u\cdot\nabla n=D_{B}{}^{2}n\label{eq:convection-gradient diffusion}
\end{equation}
where $n(\ve x$,$t$) is the particle concentration at position $\ve x$
and $D_{B}$ is the Brownian diffusion coefficient given 
\begin{equation}
D_{B}=\frac{C_{S}k_{B}T}{\eta}\label{eq:Brownian diffusion coefficient}
\end{equation}
where $\eta$ is the friction coefficient , $C_{S}$ is the slip factor
given in Eq.(\ref{eq:Cunningham slip factor}) , and $k_{B}$ is Boltzmann's
constant. We note that for a spherical particle of diameter $d_{p}$
moving in fluid of dynamic viscosity $\mu_{f}$ 
\begin{equation}
\eta=3\pi\mu_{f}d_{p}\label{eq:friction coefficient}
\end{equation}

The simplest formula of the fiber efficiency would be to assume that
the flux of particles to the surface of the fiber would be $2rD_{B}n_{0}/(r_{f}-r_{p})$
per unit length of the fiber compared with a flux per unit equivalent
area of $n_{0}U_{0}2r$ per unit length. So the efficiency of fiber
taking account of interception would be 
\begin{equation}
\varepsilon_{DR}\sim\frac{2rD_{B}}{\left(r_{f}-r_{p}\right)}\frac{1}{U_{0}2r}=\frac{D_{B}}{(r_{f}-r_{p)}U_{0}}=2Pe^{-1}(1-R)^{-1}\label{eq:gradient diffusion for Brownian diffusion}
\end{equation}
where $R$ is the interception radius $r_{p}/r_{f}$ and $Pe$ the
Peclet number $Pe=U_{0}/d_{f}D_{B}$. For interception alone, a detailed
calculation due to Lee and Liu (1982) gives 
\begin{equation}
\varepsilon_{I}=\left(\frac{1-\alpha_{f}}{Ku}\right)\left(\frac{R^{2}}{1+R}\right)\label{eq:interception capture efficiency}
\end{equation}
from which it follows that a finer fiber is more efficient than a
coarser fiber in intercepting particles. Using the Kuwabara flow field,
Eq. \ref{eq:interception capture efficiency} applies only to interception
from continuum flow around a fiber. A more detailed calculation based
on convection diffusion alone by Lee and Liu (1982) using a Kuwabara
flow field to account for a finite volume fraction for the fibers,
gives 
\begin{equation}
\varepsilon_{D}=2.6\left(\frac{1-\alpha_{f}}{Ku}\right)^{1/3}Pe^{-2/3}\label{eq:capture efficiency for convection diffusion}
\end{equation}
To account for the combined effect of interception as in the simple
Eq.\ref{eq:gradient diffusion for Brownian diffusion}, Lee and Liu
(1982) obtained an empirical formula based the sum of $\varepsilon_{I}$
and $\varepsilon_{D}$ except that the numerical coefficients 1 and
2.6 are replaced by 0.6 and 1.6, respectively. The experimental data
covered parameter values: $0.0086<\alpha_{f}<0.151,0.05<dp<1.3\mu m,\:0.0045<R<0.12$,
and $Stk<0.22$. The formula applies to submicron size particles where
there is no slip, meaning the formula is in appropriate for ufps $<100nm$
$(0.1microns)$. For ultrafine particles, which have higher diffusion
coefficients and hence smaller $Pe$, (Wang, Chen and Piu 2009) obtained
the following empirical formula: 
\begin{equation}
\varepsilon_{D}(ufps)=0.84Pe^{-0.43}\label{eq:Wang et al.}
\end{equation}
The formula in Eq.\ref{eq:Wang et al.} demonstrates a less sensitive
influence of single fiber efficiency on $Pe$ than does that given
in Eq. \ref{eq:capture efficiency for convection diffusion}. Wang,
Chen and Piu (2009) have shown that values given by the formula agree
well with the penetrations reported by Japuntich et al. (2007) and
Kim, Harrington and Pui (2007) . The data were obtained using three
face velocities, 0.053, 0.1, and 0.15 m/s, and four standard filter
media, which had effective fiber diameters of 1.9, 2.9, 3.3, and 4.9
$\mu m$ respectively, and a volume fraction $\alpha_{f}$ between
0.039 and 0.05. Subsequently, Yamada, Seto and Otani (2011) used a
glass fiber filter and a polypropylene fiber filter to study the effect
of filter structure. They were able to explian lower dependence on
$Pe$ by the non-uniformity in filter structure. Taking account the
combined effect of interception for the single fiber efficiency, Stechkina
and Fuchs (1966) used the Kuwabara flow field to obtain the following
equation: 
\begin{equation}
\varepsilon_{DI}=2.9Ku^{-1/3}Pe^{-2/3}+0.624Pe^{-1}+\varepsilon_{I}+\frac{1.24R^{2/3}}{\left(KuPe\right)^{1/2}}\label{eq:Stechkina and Fuchs}
\end{equation}
\#valid for $R<0.5,$$KuPe<0.024$..

\subsection*{Simultaneous inertial convection and Brownian diffusion}

This describes the situation where Brownian diffusion and inertial
impaction can have an equal role. It is not the case that fiber efficiency
is the sum of the two separate contributions. That we can do so is
because of the dependence of either process on particle size, when
one contribution is small, the other is large by comparison. The process
is however distinctly non linear and is treated via the underlying
Fokker-Planck equation (Drossinos and Reeks 2005) which describes
the way the particle phase space concentration $P(\vecx,\vecv,t)$
evolves with time $t$ for particles with velocity $\vecv$ and position
$\vecx$ at time $t$. The convective flux is described by an inertial
convective flux term $\tau_{p}^{-1}(\ve U_{f}$-$\vecv$) and a gradient
diffusive flux $-D_{B}\tau_{p}^{-2}\partial P/\partial v$ where $\tau_{p}$
is the Stokes relaxation time. The dispersion is described by a set
of mass momentum and energy equations where the relative contribution
of Brownian dispersion and inertial impaction is controlled by the
particle Stoke no $St.$ See Drossinos and Reeks (2005) for analytic
solutions for the case of particle dispersion in a simple shear.

\begin{figure}[H]
\begin{centering}
\includegraphics[scale=0.5]{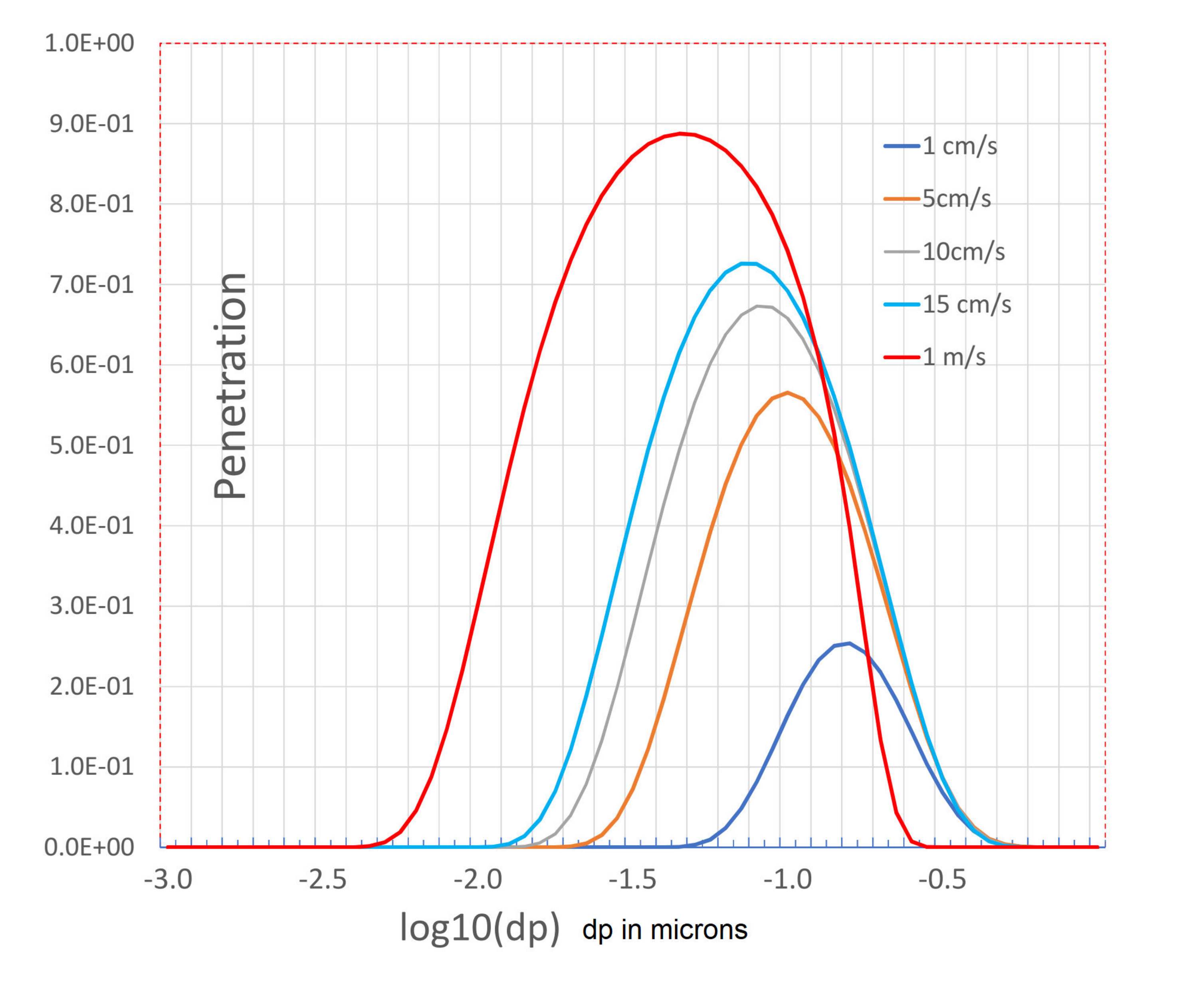} 
\par\end{centering}
\centering{}\caption{\label{fig:Fibre-filter-penetration}fiber filter penetration $p$
v particle size $(d_{p}\:microns)$ for a range of flows; $p=1$ means$100\%$penetration}
\end{figure}

\section{Calculation of penetration of fiber filters\label{sec:Calculation-of-penetration}}

In this section we present the results of the calculation of the penetration
of fiber filters based on the empirical formulae for the various components
of the fiber efficiency $\varepsilon$ given in Section \ref{sec:Filtration-Mechanisms}.
Similar values were used for the filter characteristics that have
been used in experiments and typical of those used by manufacturers
referred to in the literature e.g. Lee et al. (1982), Japuntich et
al. (2007), Yamada et al. (2011). See also the fiber characteristics
used by C-sen Wang and Y. Otani (2013): fiber diameter $df=1.7\text{\ensuremath{\mu}}m$,
and a filter thickness $L=1mm,$ with a fiber volume fraction $\alpha_{f}$
=0.07

\begin{figure}[H]
\begin{centering}
\includegraphics[scale=0.3]{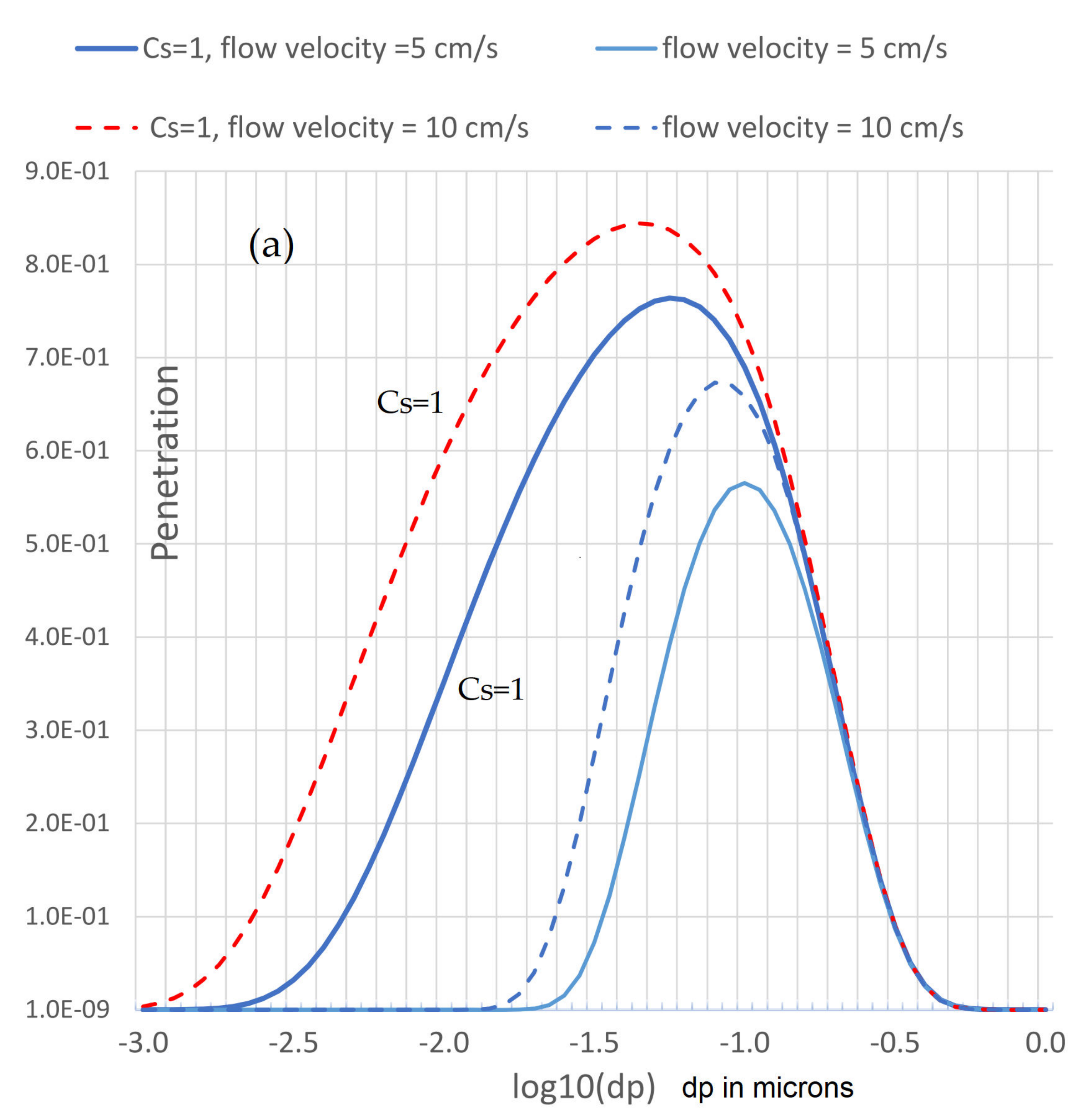} 
\par\end{centering}
\centering{}\includegraphics[scale=0.3]{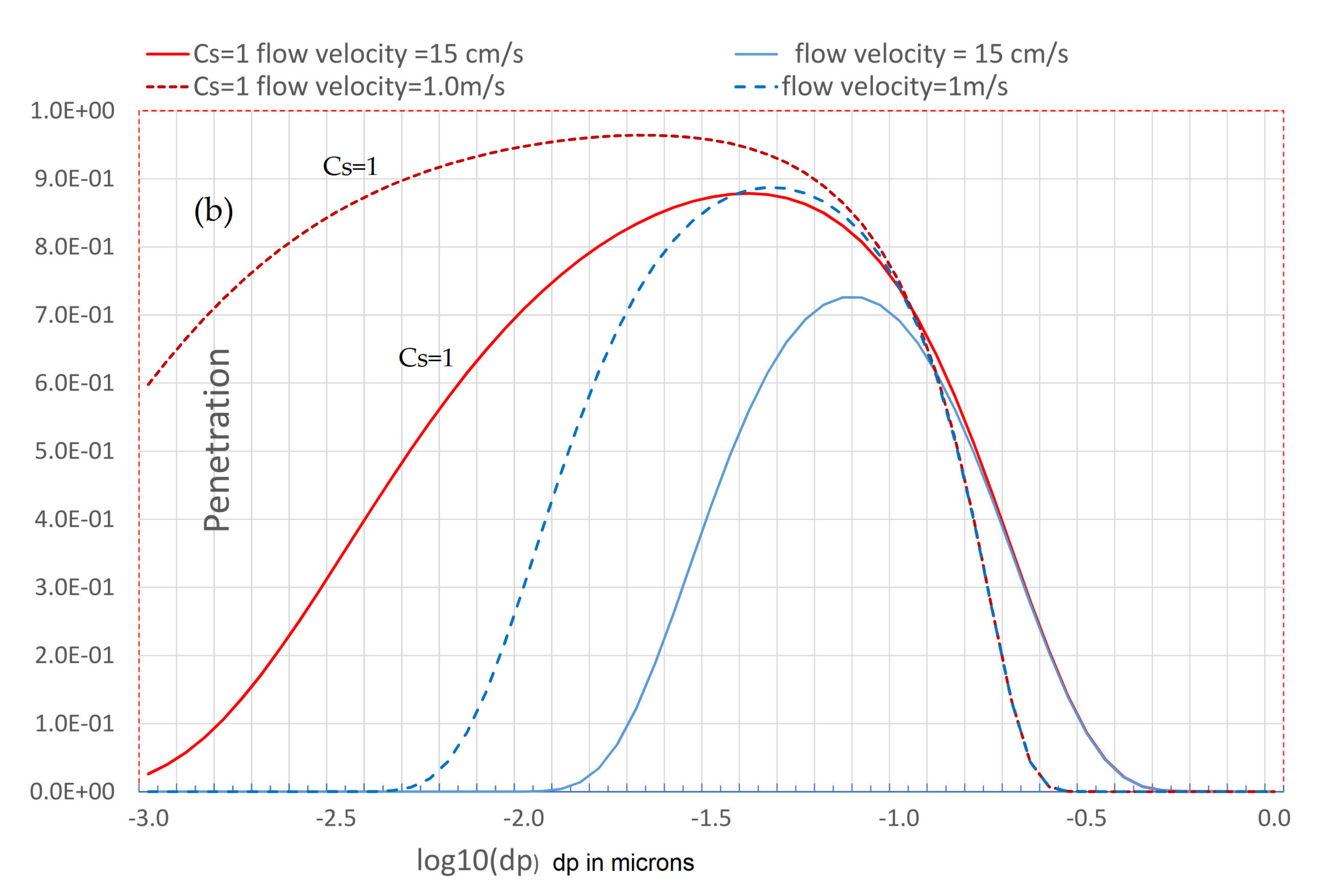}\caption{\label{fig:Penetration-for-non-slip-slip_bcs_v_flow}Penetration for
non-slip ($Cs=1$) and slip flow ($Cs=Eq.(\ref{eq:Cunningham slip factor})$}
\end{figure}

The formula used for the efficiency is based on the formula of Stechkina
and Fuchs (1966) Eq.(\ref{eq:Stechkina and Fuchs}) which accounts
for the influence of Peclet number $Pe$ and the interception ratio
$R=d_{p}/d_{f}$ and fiber volume fraction $\alpha_{f}$ on the in
fiber flow field that depends on the Kuwabara number $Ku$. To account
for the contribution to the efficiency from particle impaction , the
formula for the impaction efficiency $\varepsilon_{Imp}$ in Eq.(\ref{eq:efficiency (Stk)}),
has simply been added to the formula for $\varepsilon_{DI}$ in Eq.(\ref{eq:Stechkina and Fuchs}).
Thus 
\begin{equation}
\varepsilon_{p}=\varepsilon_{DI}+\varepsilon_{Imp}\label{eq:fibre_efffficiency}
\end{equation}

This value is then converted to a value for the penetration using
the formula given in Eq.(\ref{eq:penetration}) for a cylindrical
fiber with its axis normal to the flow, of diameter $d_{f}$ in a
filter of thickness $L$ with a fiber volume fraction $\alpha_{f}$.\\

Figures \ref{fig:Fibre-filter-penetration}-\ref{fig:Penetration-for-non-slip-slip_bcs_v_flow}
show the filter penetration as a continuous function of particle size
$d_{p}$ from $10^{-3}-1$$\mu m$ for a range of flow rates from
$1cm/s$ to $10m/s$ with both slip and no slip boundary conditions
($Cs$ =1). Figure \ref{fig:Fibre-filter-penetration} shows the penetration
as a continuous function of particle size for the complete range of
flows used in the calculation. The most important feature is a feature
common to all the graphs involving both slip and no slip bcs, namely
a maximum in the penetration which increases with flow rate so that
for $1cm/s$ $p_{m}\sim25\%$ to $p_{m}\sim90\%$. The particle size
$d_{m}$ also varies with flow rate $d_{m}\sim10^{-0.7}\mu m$ at
1 $cm/s$ to $10^{-1.3}$ for $1m/s.$ The 1/2 width of penetration
spans a 0.6 decade for 1cm/s to a 1.6 decade for $1m/s$.

\begin{figure}[H]
\centering{}\includegraphics[scale=0.55]{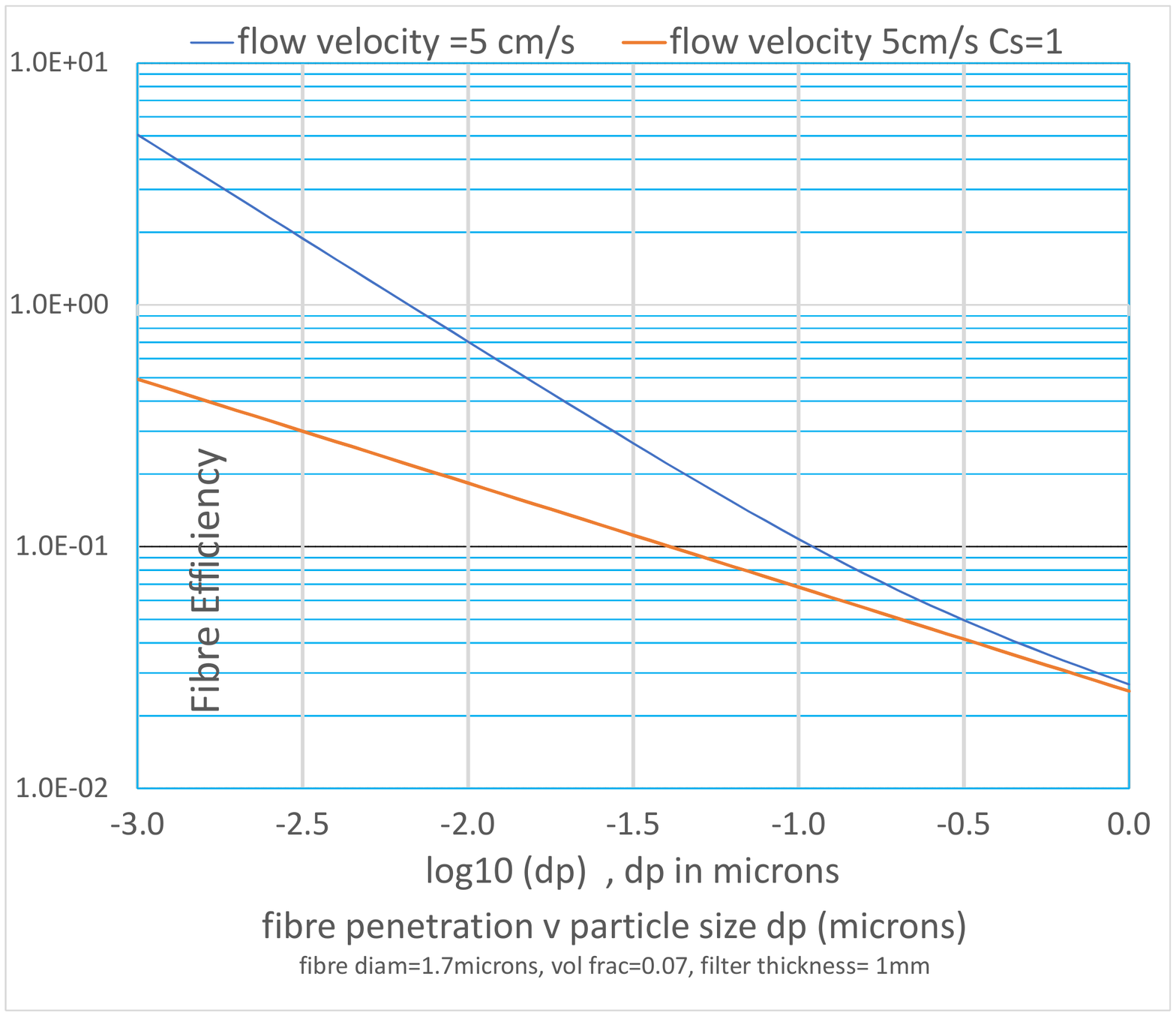}\caption{\label{fig:Fibre-efficiency-slip v no slip}fiber efficiency v particle
size $dp$ , slip / no-slip conditions $(Cs=1)$}
\end{figure}

The occurrence of a maximum penetration can be simply explained as
follows. The penetration depends essentially upon 2 numbers, the particle
Peclet number $Pe$ and the Stokes number $St$ and it is on the way
these 2 numbers depend individually upon flow and particle size that
determines how the efficiency depends overall on flow and particle
size. $Pe$ is the ratio $u_{f\,}d_{f}/D_{B}$ where $u_{f}$ is the
filter face velocity and $D_{B}$ is the Brownian diffusion coefficient.
It measures the influence of convection over diffusion. As $u_{f}$
increases so does $Pe$ and in turn the penetration $p$. Increases
in $D_{B}$ causes the value of $Pe$ to decrease and in turn to a
decrease in penetration $p.$ As far as the dependence of $p$ on
particle size, and its dependence on $Pe$, this is reflected in the
value of $D_{B}$ which decreases as particle size increases. This
causes $Pe$ to increase and $p$ in turn to increase. The other number,
namely the Stokes number $St$, increases with particle size and this
reduces the penetration. So when we consider the overall dependence
of $p$ on particle size we see that in the nano-size range $\leq.1\mu m$,
the penetration increases and for particle sizes$\geq.1\mu m$ $p$
decreases, implying a maximum value of $p$ round about $.1\mu m$.
It is to be expected that this is a general feature of the performance
of fiber filters.

We recall that the particle Brownian diffusion coefficient $D_{B}$
depends upon whether the flow around the individual particles is continuum
or non continuum, i.e. whether the boundary conditions are slip or
non slip. This depends upon the Knudsen number $Kn$, the ratio $2\lambda_{g}/d_{p}$
where $\lambda_{g}$ is the mean free path of gas molecules which
in turn determines the Cunningham slip factor $Cs$ given in Eq.(\ref{eq:Cunningham slip factor})
which modifies the Brownian diffusion coefficient from its continuum
form in Eq.(\ref{eq:Brownian diffusion coefficient}) indicating that
the diffusion coefficient under slip conditions is greater than its
continuum value for $Cs=1.$ The influence of slip over no slip bcs
on the penetration and its dependence on particle size is shown in
Figures \ref{fig:Penetration-for-non-slip-slip_bcs_v_flow} (a) and
(b). What we see is a reduction in the width of the region in which
the penetration $p$ is significant for slip over no slip bcs as well
as a reduction in $p$. The difference between the two conditions
is more marked as the particle size decreases as indicated in Fig.
\ref{fig:Fibre-efficiency-slip v no slip}

\section{Summary \& conclusions}

\noindent This paper is concerned with the release rate of ultra fine
particles (ufps) from MSW incinerators and their retention by bag/fiber
filters. It highlights the importance of using measurements and regulations
based on particle number rather than particle mass distribution versus
size for particle emissions. In particular, converting figures for
the annual particulate mass released to numbers of particle for a
number of UK incinerators, would indicate that nearly all$\geq$ 99
\% of the emitted particles are ufps with a corresponding emission
rate $\sim10^{14}$particles /s. A similar result is true of US incinerators
based on emission factors provided by the US EPA . We show how this
figure is arrived at and how it depends upon the number- size distribution
and spread of the ufps. A log normal distribution suggests that this
figure could be as high as $10^{15}$particles/s, increasing linearly
as the logarithmic spread $\sigma$ of the distribution for $\sigma\gtrsim4$.
This would imply that the bag/fiber filters have a very low efficiency
for the removal of ufps, in contrast to the assertion made by Jones
and Harrison (2016) that bag filters have a high removal efficiency
for ufps over the entire size range. An analysis of fiber filter retention
based on the fundamental mechanisms of deposition of small particles
to cylindrical fibers and their dependence on flow and particle size,
shows that whilst the removal efficiency is 100\% for particles <\textcompwordmark <
. 1 micron in size, there is a minimum of the filter retention efficiency
in the region 0.05 - 0.5 microns in size where the concentration of
the ufps is likely to be greatest. In some cases depending on the
flow and particle size, the filter efficiency is as low as 5\% compared
to almost 100\% retention efficiency for particle sizes $\gtrsim$1
micron (the inertial impaction range of particles). This behaviour
has been confirmed by accurate measurements of fiber filter penetration
as a function of individual particle size in a number of experiments.
We believe this explains the very high release rates $\sim10^{14}$particles
/ s released from these UK incinerators.

\section*{Acknowledgment}

I would like to acknowledge the support and advice of members of the
Parliamentary Particle Research Group set up by Dr. David Drew former
Shadow Minister for DEFRA, especially the convenor Ron Bailey, Shlomo
Dowen (UKWIN) and Prof. Vyvyian Howard. Finally I would also like
to thank Tim Hill and Chris Harmer (Waste Research) for their advice
and knowledge of the operation of UK incinerators and for providing
me with important data on emission factors for MSW incineration.\\

\section*{References}
\noindent \begin{flushleft}
Bowden, T. 2019. The UK will burn more than half its rubbish as it
doubles the number of incinerators over next 10 years. https://inews.co.uk/news/environment/waste-incinerators-double-burning-rubbish-air-pollution-uk-159516.\medskip{}
 \\
 Drossinos, Y. and M. W. Reeks. 2005. Brownian motion of finite-inertia
particles in a simple shear. Phys. Rev. E. 71: 031113.\smallskip{}
 \\
 Ghosh, R., A. Freni-Sterrantino, B. Parkes, D. Fecht, K. de Hoogh,
G. Fuller, J. Gulliver, A. Font, R. Smith, M. Blangiardo, P. Elliott,
M. Toledano and A. Hansell. Fetal growth, still birth infant mortality,
and other birth outcomes near municipal waste incinerators; retrospective
population based cohort and case-control study. 2019. J. Env. Int.
122:151-158.\smallskip{}
 \\
 Harrison, R. ed. 2017. Source Apportionment of Airborne Particulate
Matter in the UK, Primary Particulate Matter, Chapter 3, section 3.2.2.3.
Table 3.7, reference Emissions Factors USEPA /AP 42, cedadocs.ceda.ac.uk.\smallskip{}
 \\
 Japuntich, D. A., L. M. Franklin, D. Y. Pui, T. H. Kuehn, S. C. Kim
and A. S. Viner. 2007. A Comparison of Two Nano-Sized Particle Air
Filtration Tests in the Diameter Range of 10 to 400 Nanometers. J.
Nanopart. Res. 9: 93-107\medskip{}
 .\\
 Jones, A. M., and R. M. Harrison. 2016. Emission of ultrafine particles
from the incineration of municipal solid waste: A review. J. Atm.
Env. 140: 519-528.\medskip{}
 \\
 Kim, S. C., M. S. Harrington and D. Y. H . Pui. 2007. Experimental
Study of Nanoparticles Penetration Through Commercial Filter Media.
J. Nanopart. Res. 9: 117-125.\medskip{}
 \\
 \textcolor{black}{Kuwabara, S. 1959. The Forces Experienced by Randomly
Distributed Parallel Circular Cylinders or Spheres in a Viscous Flow
at Small Reynolds Numbers. J. Phys. Soc. Japan. 14: 527-532.}\medskip{}
 \\
 Landahl, H. and R. G. Hermann. 1949. Sampling of liquid aerosols
by wires, cylinders, and slides, and the efficiency of impaction of
the droplets. J. Colloid. 4: 103-136.\medskip{}
 \\
 Lee, K. W. and B. Y. H. Liu. 1982. Theoretical Study of Aerosol Filtration
by Fibrous Filters. Aerosol Sci. Technol. 1: 147-161.\smallskip{}
 \\
 NAEI data base. 2016. Particulate annual emissions and mass-size
distribution from UK Incinerators, Air Pollution Emissions reports.\smallskip{}
 \\
 Nixon, J. D., D. G. Wright, P. K. Deyb, K. Ghosh, and P. A. Davies.
2013. A comparative assessment of waste incinerators in the UK. Waste
Management. 33 (11): 2234-2244.\smallskip{}
 \\
 Oberdorster, G. 2001. The pulmonary effects of inhaled ultrafine
particles. Int Arch Occup Environ. Health. 74:1-8.\smallskip{}
 \\
 Pich, J. 1966. Theory of Aerosol Filtration by Fibrous and Membrane
Filters. In Aerosol Science ed. C. N. Davies, 223-280. Academic Press:
London.\medskip{}
 \\
 Shang, Yu., Wu Meiying, Jizhi Zhou, Yufang Zhong, Jing An, and Guangren
Quian. 2019. Cytotoxicity comparison between fine particles emitted
from the combustion of municipal solid waste and biomass. J. Hazardous
Materials. 367: 316--324.\medskip{}
 \\
 Stechkina, I. B., and N. A. Fuchs. 1966. Studies on Fibrous Aerosol
Filters\textminus I. Calculation of Diffusional Deposition of Aerosols
in Fibrous Filters. Ann. Occup. Health. 9: 59-64.\medskip{}
 \\
 Wang, C-sen., and Y. Otani. 2013. Removal of Nanoparticles from Gas
Streams by Fibrous Filters: A Review. Ind. Eng. Chem. Res. 52: 5\textminus 17.\medskip{}
 \\
 Wang, J., D. R. Chen and D. Y. H. Pui. 2007. Modeling of Filtration
Efficiency of Nanoparticles in Standard Filter Media. J. Nanopart.
Res. 9: 109-115.\medskip{}
 \\
 Yamada, S., T. Seto and Y. Otani. 2011. Influence of Filter Inhomogeneity
on Air Filtration of Nanoparticles. Aerosol Air Qual. Res. 11: 155-160.\\
 
\par\end{flushleft}
\end{document}